\documentclass[12pt]{article}

\begin{document}
\thispagestyle{empty}
\begin{center}
{\Large \bf Analytical five-loop expressions for
the renormalization group  QED  $\beta$-function  in different 
renormalization schemes}

\vspace{0.1cm}
{A.~L.~Kataev and S.A.Larin}\\
\vspace{0.1cm}

Institute for Nuclear
Research of the Russian Academy of Sciences,\\  
117312 Moscow, Russia
\end{center}
\begin{abstract}
We obtain analytical five-loop results for the renormalization group $\beta$-function of Quantum Electrodynamics with
the single lepton in different renormalization schemes. The theoretical consequences of the results obtained are
discussed.
\end{abstract}
PACS numbers: 03.70.+k; 11.10.-z; 11.15.Bt\\

The version of the text, published in Pisma v ZhETF, vol.96, iss.1, pp.64-67.

\newpage 
The concept of the   $\beta$-function, which depends on the choice of  the  renormalization scheme,   is    the   cornerstone
of the Quantum Field  Theory  renormalization group approach, developed in the works
of Refs.\cite{Stueckelberg:1953dz},\cite{GellMann:1954fq},\cite{Bogolyubov:1956gh}.
In QED the study of the  perturbative expansion of the  $\beta$-function
is of special interest. Indeed, it governs the  energy-dependence of  the  constant   $\alpha=e^2/(4\pi)$, which defines
the coupling of  photons with
leptons. In this work we will obtain  the five -loop analytical expressions for  the renormalization group QED
$\beta$-function of the  electron, neglecting
the contributions of leptons with higher masses, namely the contributions of  muons and tau-leptons.

We will start with the expression for the  $\beta$-function in the
variant of the  minimal subtraction
scheme \cite{'tHooft:1973mm} , namely  the
$\overline{MS}$-scheme \cite{Bardeen:1978yd}:
\begin{equation}
\beta_{\overline{MS}}(\bar{\alpha})=\mu^2\frac{\partial ({\bar{\alpha}/\pi})}{\partial \mu^2}
=\sum_{i\geq 1} {\bar\beta}_{i} \bigg(\frac{\bar{\alpha}}{\pi}\bigg)^{i+1}
\end{equation}
where $\bar{\alpha}$ is the renormalized  $\overline{MS}$-scheme  QED coupling constant and $\mu^2$ is the $\overline{MS}$-scale parameter.
At the three -loop level the scheme-dependent  coefficient ${\bar\beta}_3$ was independently  evaluated analytically in Ref.\cite{Vladimirov:1979zm}
and Ref.\cite{Chetyrkin:1980pr}. The four-loop coefficient  ${\bar\beta}_4$ was obtained as the result of the  project, started in
Ref.\cite{Gorishnii:1987fy}
and completed in  Ref.\cite{Gorishnii:1990kd}. The QED  result of Ref. \cite{Gorishnii:1990kd} was confirmed  after taking
the  QED limit of the analytically evaluated
in Ref. \cite{vanRitbergen:1997va} 4-loop correction to the
$\overline{MS}$-scheme $\beta$-function of the $SU(N_c)$ colour gauge model.

To get the five-loop expression for the coefficient $\bar{\beta}_5$ we use the derived in Ref. \cite{Nigam:1994yz}  renormalization-group expression,
which has the following form:
\begin{equation}
\label{beta5}
\bar{\beta}_5=-\bar{b}_5-3\bar{b}_1\bar{a}_4-2\bar{b}_2\bar{a}_3- \bar{b}_3\bar{a}_2-\bar{b}_1\bar{a}_2^2
\end{equation}
where $\bar{a}_l$ and $\bar{b}_l$ enter into the expressions for
the l-loop  contributions to the  photon vacuum polarization
functions with  $1\leq l \leq 5$. These contributions  are defined as
\begin{equation}
\label{Pi}
\bar{\Pi}_l(x)=(\bar{a}_l+\bar{b}_l ln(x)+\bar{c}_l ln^2(x)+\bar{d}_l ln^3(x)+\bar{e}_l
ln^4(x))\bigg(\frac{\bar{\alpha}}{\pi}\bigg)^{l-1}
\end{equation}
where $x=Q^2/\mu^2$ and $Q^2$ is the Euclidean momentum transfer.   It is possible to show, that for
 $l\leq 2$  $\bar{c}_l=0$, $\bar{d}_l=0$, $\bar{e}_l=0$, for $l\leq 3$ $\bar{d}_l=0$ and $\bar{e}_l=0$, while for $l=4$ $\bar{e}_4=0$. In general
$\bar{c}_l$,$\bar{d}_l$, $\bar{e}_l$ are expressed through the products of  lower order coefficients $\bar{b}_i$ via the corresponding renormalization group relations
(see Refs.\cite{Gorishnii:1991hw},\cite{Nigam:1994yz}). For $1 \leq l \leq 3$ the coefficients $\bar{a}_i$ and $\bar{b}_i$ are defined by the  the results
of Ref.\cite{Gorishnii:1991hw}
and read $\bar{a}_1=5/9$, $\bar{a}_2=55/48 -\zeta_3$, $\bar{a}_3=-1247/648-(35/72)\zeta_3+(5/2)\zeta_5$,  $\bar{b}_1=-1/3$, $\bar{b}_2=-1/4$,
$\bar{b}_3=47/96-\zeta_3/3$, while the expression for $\bar{a}_4=1075825/373248-(13/96)\zeta_4+(13051/2592)\zeta_3-(5/3)\zeta_3^2+
(45/32)\zeta_5-
(35/4)\zeta_7$ was evaluated  in Ref.\cite{Baikov:2008si}. In order to get  $\bar{\beta}_5$ from Eq.(\ref{beta5})
we should fix the analytical expression for $\bar{b}_5$ in the  r.h.s. of Eq.(\ref{beta5}).
It is composed of  the sum of three  contributions
$\bar{b}_5=\bar{b}_5[nlbl]+\bar{b}_5[3,lbl]+\bar{b}_5[2,lbl]$. The first  term in $\bar{b}_5$ is fixed by the leading logarithmic term of
the sum of    five-loop photon vacuum polarization   graphs, which contain
the  electron  loop with two external vertexes.  The second  term  $\bar{b}_5[3,lbl]$ comes  from the   logarithmic contribution  to five-loop
photon propagator diagrams with two one-loop  light-by-light scattering type sub-graphs, connected by a  photon line with an electron loop
insertion and two undressed photon propagators.
The third  term $\bar{b}_5[2,lbl]$ arises from the  logarithmic contribution  to five-loop
photon propagator diagrams with  one-loop  and two-loop  light-by-light scattering type sub-graphs,  connected by  three undressed photon
propagators.

The first contribution into $\bar{b}_5$ can be   defined from the QED limit of the result of Ref.\cite{Baikov:2010je}
with one active lepton  and reads

\begin{equation}
\label{1}
\bar{b}_5[nlbl]=-\frac{409367}{1492992}+\frac{15535}{2592}\zeta_3-\zeta_3^2+\frac{25}{12}\zeta_5-\frac{35}{4}\zeta_7
\end{equation}

The term $\bar{b}_5[3,lbl]$ is fixed by us from   the  QED limit of the   depending on the  number of fermions analytical
expression for the order $\alpha_s^4$ singlet QCD  contribution to the cross-section for electron-positron annihilation
into hadrons, published    in Ref.\cite{Baikov:2011POS}. It has the following form
\begin{equation}
\label{2}
\bar{b}_5[3,lbl]=\frac{149}{108}-\frac{13}{6}\zeta_3-\frac{2}{3}\zeta_3^2+\frac{5}{3}\zeta_5
\end{equation}

The term $\bar{b}_5[2,lbl]$ is obtained from the QED limit of the $C_F\alpha_s^4$ singlet QCD contribution to the
cross-section for electron-positron annihilation
into hadrons, presented in the subsequent  works of Refs.\cite{Baikov:2010iw}, \cite{Kataev:2011im}, with $C_F$ being
the quadratic Casimir operator of the $SU(N_c)$ colour gauge group.  This term is
\begin{equation}
\label{3}
\bar{b}_5[2,lbl]=\frac{13}{12}+\frac{4}{3}\zeta_3-\frac{10}{3}\zeta_5
\end{equation}

Substituting the results from Eq.(\ref{1}), Eq.(\ref{2}) and Eq.(\ref{3}) into the r.h.s. of Eq.(\ref{beta5}) we get the analytical
expression for the five-loop approximation of  the $\overline{MS}$-scheme QED  $\beta$-function
with a  single lepton:
\begin{eqnarray}
\beta_{\overline{MS}}(\bar{\alpha})&=&\mu^2\frac{\partial ({\bar{\alpha}/\pi})}{\partial \mu^2}
=\sum_{i\geq 1} {\bar\beta}_i \bigg(\frac{\bar{\alpha}}{\pi}\bigg)^{i+1} \nonumber  \\
&=&\frac{1}{3}\bigg(\frac{\bar{\alpha}}{\pi}\bigg)^2+\frac{1}{4}\bigg(\frac{\bar{\alpha}}{\pi}\bigg)^3-
\frac{31}{288}\bigg(\frac{\bar{\alpha}}{\pi}\bigg)^4-\bigg(\frac{2785}{31104}+\frac{13}{36}\zeta_3\bigg)\bigg(\frac{\bar{\alpha}}
{\pi}\bigg)^5 \nonumber  \\ \label{5loop}
&+&\bigg(-\frac{195067}{497664}-\frac{13}{96}\zeta_4-\frac{25}{96}\zeta_3+\frac{215}{96}\zeta_5\bigg)
\bigg(\frac{\bar{\alpha}}
{\pi}\bigg)^6 +O(\bar{\alpha}^7)
\end{eqnarray}
which contain the contributions of the Riemann $\zeta$-functions, defined as
$\zeta_{\rm k}=\sum_{n=1}^{\infty}(1/n)^{\rm k}$.
Let us remind that  scheme-dependent  coefficients of the $\beta$-function  do not depend on the concrete realization of the minimal subtraction scheme
(see e.g. \cite{Chetyrkin:1980pr}). Notice the appearance of the $\zeta_4$-term in the expression
for $\bar{\beta}_5$, which did not manifest itself in the lower order coefficients.
This feature was already observed in
Ref.\cite{Baikov:2001aa}
as the result of analytical calculations of the cubic in the number of leptons
five-loop terms of  the the QED  $\beta_{\overline{MS}}$ -function, which did not contain
the cubic in the number of leptons  light-by-light-type terms. Comparing our  result of Eq.(\ref{5loop}) with
the  expression from Ref.\cite{Baikov:2001aa}, we conclude that the addition of the
5-loop light-by-light-type contributions changes the coefficient  and sign of the  $\zeta_4$-contribution in the
overall expression for $\bar{\beta}_5$, given in Eq.(\ref{5loop}). This happens due to taking into account in the second
term of Eq.(\ref{beta5})  the light-by-light-type contribution into the constant term $\bar{a}_4$.
Another intriguing observation is the cancellation in  Eq.(\ref{5loop})
of the $\zeta_7$ and $\zeta_3^2$ transcendentalities, which contribute to the first term in Eq.(\ref{beta5}), namely
the $\bar{b}_5[nlbl]$-term (see Eq.(\ref{1})).

Let us now transform Eq.(\ref{beta5}) from the $\overline{MS}$-  to the on-shell scheme using the following
equation
\begin{equation}
\label{betaos}
\beta_{OS}(\alpha)=\sum_{i\geq 1}\beta_i\bigg(\frac{\alpha}{\pi}\bigg)^{i+1}=\beta_{\overline{MS}}(\bar{\alpha}(\alpha))/{\frac{\partial \bar{\alpha}(\alpha)}{\partial \alpha}}
\end{equation}
where  $\mu^2=m^2$, $m$ is the electron  pole mass, $\alpha$ is the QED coupling
constant, defined in the on-shell scheme   and
\begin{equation}
\label{transform}
\bar{\alpha}(\alpha)=\alpha\bigg(1+g_2\bigg(\frac{\alpha}{\pi}\bigg)^2+g_3\bigg(\frac{\alpha}{\pi}\bigg)^3+g_4
\bigg(\frac{\alpha}{\pi}\bigg)^4+O(\alpha^5)\bigg)~~~~.
\end{equation}
The coefficients    $g_2=15/16$, $g_3=-4867/5184+(23/72)\pi^2-(1/3)\pi^2{\rm ln}2
+(11/96)\zeta_3$ were evaluated in Ref.\cite{Broadhurst:1992za}, while
 $g_4=14327767/9331200+ (8791/3240)\pi^2+(204631/259200)\pi^4-(175949/4800)\zeta_3+(1/24)\pi^2\zeta_3
+(9887/480)\zeta_5-(595/108)\pi^2 {\rm ln}2 -(106/675)\pi^4{\rm ln}2 +(6121/2160)\pi^2{\rm ln}^2~2- (32/135)\pi^2{\rm ln}^3~2-
(6121/2160){\rm ln}^4~2+(32/225){\rm ln}^5~2-(6121/90)a_4-(256/15)a_5$ with  $a_4$ and $a_5$
defined  as  $a_{\rm k}=\rm {Li_k[1/2]}=\sum_{n=1}^{\infty}(1/2n)^k$ was obtained in Ref.\cite{Baikov:2008si}.

Using these results in the transformation relations of Eq.(\ref{transform}) and Eq.(\ref{betaos}) we get
\begin{eqnarray}
\beta_{OS}(\alpha)&=&m^2\frac{\partial ( \alpha/\pi)}{\partial m^2}
=\sum_{i\geq 1} \beta_i \bigg(\frac{\alpha}{\pi}\bigg)^{i+1} \nonumber  \\
&=&\frac{1}{3}\bigg(\frac{\alpha}{\pi}\bigg)^2+\frac{1}{4}\bigg(\frac{\alpha}{\pi}\bigg)^3
-\frac{121}{288}\bigg(\frac{\alpha}{\pi}\bigg)^4
+\bigg(\frac{5561}{10368}-\frac{23}{18}\zeta_2+\frac{4}{3}\zeta_2 {\rm ln}2-\frac{7}{16}\zeta_3
\bigg)\bigg(\frac{\alpha}
{\pi}\bigg)^5 \nonumber  \\
&+&\bigg(-\frac{23206993}{37324800}+
\frac{6121}{2160}{\rm ln}^4 2 -\frac{32}{225}{\rm ln}^5 2 -\frac{205021}{259200}\pi^4
+\frac{106}{675}\pi^4{\rm ln} 2 +\frac{6121}{90}{\rm Li}_4(1/2) \nonumber  \\
&+&\frac{256}{15}{\rm Li}_5(1/2)
- \frac{36199}{12960}\pi^2+\frac{151}{27}\pi^2{\rm ln} 2  -\frac{6121}{2160}\pi^2 {\rm ln}^2 2
+\frac{32}{135}\pi^2 {\rm ln}^3 2 \\ \nonumber
&-&\frac{1}{24}\pi^2 \zeta_3+\frac{349123}{9600}\zeta_3
-\frac{2203}{120}\zeta_5\bigg)\bigg(\frac{\alpha}
{\pi}\bigg)^6 +O(\alpha^7)
\end{eqnarray}
The third coefficient coincides with the one, originally calculated in Ref.\cite{DeRafael:1974iv}.
The agreement between   analytical
results for $\bar{\beta}_3$ and $\beta_3$-coefficients was first demonstrated in
Ref.\cite{Chetyrkin:1980pr}. The expression for
the four-loop coefficient $\beta_4$ is in agreement with the result of Ref.\cite{Broadhurst:1992za}.
The five-loop coefficient $\beta_5$ is new. Note, that both $\beta_4$ and $\beta_5$-terms contain typical
to the on-shell renormalization procedure contributions, which are proportional to $\rm{ln}2$ and
$\zeta_2=\pi^2/6$. However, at present  we are unable to rewrite the
proportional to $\pi^{2}$ contributions into $\beta_5$ through the
$\zeta$-functions of even arguments. Indeed, the  $\pi^4$-contributions to  $g_4$ may be decomposed
into the sum of $\zeta_4$ and $\zeta_2^2$-terms with unknown to us coefficients.

In order to get the five-loop expression for the QED  Gell-Mann-Low  $\Psi(\tilde{\alpha})$-function, which
coincides with the QED $\beta$-function in the momentum (MOM) subtractions scheme (for the  detailed explanation of
this statement at the four-loop level   see Refs.\cite{Gorishnii:1987fy},\cite{Gorishnii:1990kd}) we supplement
the general  transformation relation  between the  $\beta_{OS}(\alpha)$-function  and the $\Psi$-function,
derived in Ref.\cite{Nigam:1994yz} with the explicit results for the on-shell scheme  analogs of the  coefficients $\bar{a}_3$ and
$\bar{a}_4$ in Eq.(\ref{Pi}), which are known from the results of Ref.\cite{Broadhurst:1992za} and Ref.\cite{Baikov:2008si}
respectively. The obtained result reads
\begin{eqnarray}
\Psi(\tilde{\alpha})&=&\mu^2\frac{\partial (\tilde{ \alpha}/\pi)}{\partial \mu^2}
=\sum_{i\geq 1} \Psi_i \bigg(\frac{\tilde{\alpha}}{\pi}\bigg)^{i+1} \nonumber  \\
&=&\frac{1}{3}\bigg(\frac{\tilde{\alpha}}{\pi}\bigg)^2+\frac{1}{4}\bigg(\frac{\tilde{\alpha}}{\pi}\bigg)^3+
\bigg(-\frac{101}{288}+\frac{1}{3}\zeta_3\bigg)\bigg(\frac{\tilde{\alpha}}{\pi}\bigg)^4+
\bigg(\frac{93}{128}+\frac{1}{3}\zeta_3-\frac{5}{3}\zeta_5\bigg)
\bigg(\frac{\tilde{\alpha}}{\pi}\bigg)^5 \nonumber \\ \label{Psi5}
&+&
\bigg(-\frac{122387}{55296}-\frac{79}{24}\zeta_3+ \zeta_3^2
-\frac{185}{72}\zeta_5+\frac{35}{4}\zeta_7 \bigg)
\bigg(\frac{\tilde{\alpha}}{\pi}\bigg)^6 +O(\tilde{\alpha}^7)
\end{eqnarray}
We checked that the identical result is obtained from the five-loop expression for the $\beta_{\overline{MS}}(\bar{\alpha})$-function
after transforming it into the MOM-scheme. The
expressions for  $\Psi_3$ and $\Psi_4$ coincide with the results,  originally obtained in Ref.\cite{Baker:1969an}
and Ref.\cite{Gorishnii:1990kd}. The expression for $\Psi_5$ is new. On the contrary to the five-loop
$\overline{MS}$- and on-shell scheme coefficients
$\bar{\beta}_5$ and $\beta_5$ it does not contain $\zeta$-functions of even arguments. However, the
contributions of  $\zeta_7$ and $\zeta_3^2$-terms  manifest themselves in $\Psi_5$ only. They are
related to the similar scheme-independent \cite{Baikov:2008si} contributions into the non-logarithmic four-loop coefficients $\bar{a}_4$ and $a_4$
of  the renormalized photon vacuum polarization function in the $\overline{MS}$- and on-shell scheme,
as given in Ref. \cite{Baikov:2008si}.

For the completeness we present the five-loop expression   for the perturbative quenched QED contribution to the QED $\beta$-functions.
It was originally obtained in Ref.\cite{Baikov:2008cp} and published  later on in  Ref.\cite{Baikov:2010je} after additional theoretical  cross-checks, proposed
in Ref.\cite{Kataev:2008sk}. The result reads
\begin{equation}
F_1(\alpha_{*})=\frac{1}{3}\bigg(\frac{\alpha_{*}}{\pi}\bigg)+\frac{1}{4}\bigg(\frac{\alpha_{*}}{\pi}\bigg)^2-
\frac{1}{32}\bigg(\frac{\alpha_{*}}{\pi}\bigg)^3-\frac{23}{128}\bigg(\frac{\alpha_{*}}{\pi}\bigg)^4
+\bigg(\frac{4157}{6144}+\frac{1}{8}\zeta_3\bigg)\bigg(\frac{\alpha_{*}}{\pi}\bigg)^5+O(\alpha_{*}^6)
\end{equation}
where $\alpha_{*}$ is the corresponding  expansion parameter, while the coefficients of $F_1$-function do not
depend from the renormalization scheme.

At the three- and four-loop level the related expressions were obtained in Ref.\cite{Rosner:1967zz}
and Ref.\cite{Gorishnii:1990kd}. The 4-loop result was independently confirmed later on in
the work of Ref.\cite{Broadhurst:1999zi}.

In the numerical form the five-loop  perturbative series we are interested in read
\begin{eqnarray}
\beta_{\overline{MS}}(\bar{\alpha})&=&0.3333 \bigg(\frac{\bar{\alpha}}{\pi}\bigg)^2+0.25\bigg(\frac{\bar{\alpha}}{\pi}\bigg)^3
-0.1076\bigg(\frac{\bar{\alpha}}{\pi}\bigg)^4-0.5236\bigg(\frac{\bar{\alpha}}{\pi}\bigg)^5+
1.471\bigg(\frac{\bar{\alpha}}{\pi}\bigg)^6 \label{betaMS}
\\
\beta_{OS}(\alpha)&=&0.3333 \bigg(\frac{\alpha}{\pi}\bigg)^2+0.25\bigg(\frac{\alpha}{\pi}\bigg)^3
-0.4201\bigg(\frac{\alpha}{\pi}\bigg)^4-0.5712\bigg(\frac{\alpha}{\pi}\bigg)^5
-0.3462\bigg(\frac{\alpha}{\pi}\bigg)^6   \label{betaOS} \\
\Psi(\tilde{\alpha})&=&0.3333 \bigg(\frac{\tilde{\alpha}}{\pi}\bigg)^2+0.25\bigg(\frac{\tilde{\alpha}}{\pi}\bigg)^3
+0.04999\bigg(\frac{\tilde{\alpha}}{\pi}\bigg)^4-0.6010\bigg(\frac{\tilde{\alpha}}{\pi}\bigg)^5+
1.434\bigg(\frac{\bar{\alpha}}{\pi}\bigg)^6   \label{Psi}\\
F_1(\alpha_{*})&=&0.3333\bigg(\frac{\alpha_{*}}{\pi}\bigg)+0.25\bigg(\frac{\alpha_{*}}{\pi}\bigg)^2
-0.03125\bigg(\frac{\alpha_{*}}{\pi}\bigg)^3-0.1797(\frac{\alpha_{*}}{\pi}\bigg)^4
+0.8268\bigg(\frac{\alpha_{*}}{\pi}\bigg)^5 \label{F1}
\end{eqnarray}

Let us discuss the structure of these perturbative series. From the theoretical  arguments, presented in the work of
Ref.\cite{Lipatov:1976ny} one may expect that the $\beta$-functions are expanded into the sign-alternating asymptotic  perturbative series with
fast  growing  coefficients.  And indeed, this feature is true in the case of  the $\beta$-function of the $g\phi^4$-
theory, which is known in the $\overline{MS}$-scheme up to the  five-loop \cite{Gorishnii:1983gp},
\cite{Kleinert:1991rg}. In the case of QED the asymptotic estimates of Ref.\cite{Itzykson:1977mf} and Ref. \cite{Bogomolny:1982ea},   analogous to Lipatov's ones
for the  $g\phi^4$-theory \cite{Lipatov:1976ny}, indicate that the asymptotic structure of  QED perturbative series is more complicated, than in the $g\phi^4$
theory. Indeed, the asymptotic   of Refs.\cite{Itzykson:1977mf},\cite{Bogomolny:1982ea} were obtained only  for the gauge-invariant subclasses of diagrams
with  fixed number of fermion loops. Moreover,   the indication of the sign-alternating factorial 
growth of  the perturbative coefficients of
the $F_1$-function, given in  Ref.\cite{Itzykson:1977mf},  does  not agree with the concrete behavior
of the five-loop perturbative series, presented in   Eq.(\ref{F1}). Besides, as it is  discussed in Ref.\cite{Itzykson:1977mf} and
 Ref. \cite{Bogomolny:1982ea}, in the case of complete QED  the  strong cancellations
 between coefficients of sub-sets of diagrams with different fixed numbers  of fermion loops is expected. This effect may  manifest
 itself in the differences of sign structures of the five-loop approximations for $\beta_{\overline{MS}}(\bar{\alpha})$,
 $\beta_{OS}(\alpha)$  and $\Psi(\tilde{\alpha})$ (compare  Eq.(\ref{betaMS}) with Eq.(\ref{betaOS}) and Eq.(\ref{Psi})).

\newpage
It is also interesting to note, that taking into account  the calculated by us five-loop correction to $\Psi(\tilde{\alpha})$
confirms the confidence in the validity of the criterion $O\leq \Psi(\tilde{\alpha})<(\tilde{\alpha}/\pi)$,  derived
by Schwinger \cite{Schwinger:1975th} and Krasnikov \cite{Krasnikov:1981rp} (see Ref.  \cite{Yamagishi:1981di} as well).
Note also, that the theoretical analysis of the behavior of the  perturbative series for the Gell-Mann-Low function
 $\Psi(\tilde{\alpha})$, preformed
in Ref.\cite{Suslov:2001vx}, which  at large $\tilde{\alpha}$   indicates the validity of its  linear  behavior, supports
 the mentioned above identity, derived in Refs.\cite{Schwinger:1975th},\cite{Krasnikov:1981rp},
\cite{Yamagishi:1981di}.

However, we think that the  the similar linear behavior, obtained  in Ref.
\cite{Suslov:2008sh} for the QED $\beta$-function in the on-shell scheme for the case when the expansion parameter is
going to infinity should be reconsidered. Indeed, analyzing the behavior of the   perturbative series 
for the 
$\beta_{OS}(\alpha)$-function  from Eq.(\ref{betaOS}) at the three- four- and five-loop levels, we observe
the appearances of the rigorously  speaking unphysical ultraviolet fixed points at $\alpha/\pi\approx 1.2 $, $\alpha/\pi \approx 0.8$
and $\alpha/\pi\approx 0.7$ respectively. The appearances of these zeros may affect the exact asymptotic behavior of
the QED $\beta$-function in the on-shell scheme, considered in Ref.
\cite{Suslov:2008sh}.

{\bf Acknowledgments}
This work was done using the  Computational cluster of the 
Theory Division of the  Institute for  Nuclear Research
of the Russian Academy of Sciences and is supported by 
the Grant NS-5590.2012.2.
The work of one of us (ALK) was done within the Scientific 
Program of RFBR Grants N 11-01-00182, and N 11-02-00112.

After the acceptance of this work for publication the analytical 
expression for the 5-loop QED corrections to the 
$\beta_{\overline{MS}}(\overline{\alpha})$ and $\Psi$-function with $N$-number 
of identical charged leptons became known  \cite{Baikov:2012zm}.

\end{document}